\documentclass[aps, pra, twocolumn, longbibliography, floatfix]{revtex4-2}
\usepackage[english]{babel}
\usepackage[utf8]{inputenc}
\usepackage[colorinlistoftodos, color=green!40, prependcaption]{todonotes}
\usepackage{amsthm}
\usepackage{mathtools}
\usepackage{physics}
\usepackage{xcolor}
\usepackage{graphicx}
\usepackage{adjustbox}
\usepackage{placeins}
\usepackage[T1]{fontenc}
\usepackage{lipsum}
\usepackage{csquotes}
\usepackage{comment}
\usepackage{subcaption}
\usepackage{ragged2e}
\usepackage{amssymb}
\usepackage{times}
\usepackage{bm}
\makeatletter
\AtBeginDocument{
    \def\@makecaption#1#2{
        \par
        \vskip\abovecaptionskip
        \begingroup
            \small     
            \justifying 
            \textbf{#1.}\ #2\par
        \endgroup
        \vskip\belowcaptionskip
        \par
    }
}
\makeatother

\usepackage[colorlinks,citecolor=blue,linkcolor=red]{hyperref}

\begin{document}
\title{Optically Switched Phonon Superradiance of Surface Acoustic Wave in Diamond}

\author{Zhiwei Chen}
\thanks{These authors contributed equally to this work.}
%\affiliation{School of Physics Science and Engineering, Tongji University, Shanghai 200092, China}

\author{Changyong Lei}
\thanks{These authors contributed equally to this work.}
%\affiliation{School of Physics Science and Engineering, Tongji University, Shanghai 200092, China}

\author{Jie Ren}
\email{xonics@tongji.edu.cn}
\affiliation{Center for Phononics and Thermal Energy Science, China-EU Joint Lab on Nanophononics, School of Physics Science and Engineering, Tongji University, Shanghai 200092, China}

\date{\today}

\begin{abstract}
Surface acoustic wave (SAW) phonon coupling with nitrogen-vacancy (NV) center spins in diamond offers a promising platform for on-chip quantum phononic manipulations. Although an ensemble of NV centers coupled to a common SAW phonon mode enables superradiance and collective quantum control, achieving a tunable superradiant phase transition remains challenging. Here, we show that optically driving NV centers level transitions enhances the effective spin-phonon coupling, triggering a SAW phonon superradiant phase transition in the weak-coupling regime.  We also demonstrate that above a critical threshold, the driving light rapidly switches on the phonon superradiance--a dynamic effect that persists in finite-number NV ensembles. Our results provide a controllable route to coherent phonon-NV spin manipulation in solid state quantum devices.

\end{abstract}

\maketitle

\section{Introduction}
The manipulation of quantum properties of phonons, including phonon coherent \cite{coherent,huangkun,YuenH}, squeezed \cite{squeeze,zhengh1,zhengh2,zhangh3}, entanglement states \cite{entangle}, and phonon spin \cite{phononspin1,phononspin2}, is of paramount importance in the field of quantum phononics \cite{quantumphononics}, with applications in quantum sensing \cite{senor}, quantum information \cite{quantuminfor} and computation \cite{quancomp}. The generation of a coherent sound oscillations--phonon laser \cite{phonlasr1,phonlasr2,phonlasr3,phonlasr4,phonlasr5} provides a variety of applications, including high-precision sensing and imaging \cite{senor}, and gravitational wave detection \cite{gravitywave}. In particular, the development of the surface acoustic wave (SAW) phonon laser \cite{phonlasr6, SAWphlaser} for the purpose of on-solid-state quantum information \cite{Chip} is of significant importance. 
However, conventional phonon laser schemes resemble optical lasers, whose two-level emitters in the optical cavity operate independently without accounting for collective effects. Superradiance \cite{PhysRev.93.99} is the phenomenon where an ensemble of $N$ emitters collectively coupled to a single-mode radiation field emits radiation with peak intensity $I \sim N^2$, considerably exceeding the $I \sim N$ scaling of conventional lasers.
%Superradiance \cite{PhysRev.93.99} is a phenomenon in which an ensemble of $N$ emitters, collectively coupled to a single-mode radiation field, coherently emits radiation of a peak intensity $I \sim N^2$. This is significantly higher than the intensity of conventional lasers, which typically yield $I \sim N$. %This is in contrast to conventional lasers, which typically yield $I \sim N$. 
Superradiant phase transition occurs when coupling exceeds a critical threshold \cite{hepp1973phase, PhysRevA.7.831}, spontaneously breaking symmetry to generate macroscopic radiation field population \cite{squeezesup}.
%There is a phase transition between normal radiation (normal phase) and superradiance (superradiant phase) due to spontaneous symmetry breaking \cite{squeezesup}. A superradiant phase transition occurs when the emitter-radiation field coupling exceeds a critical threshold \cite{hepp1973phase, PhysRevA.7.831}, in which the radiation field becomes macroscopically populated (spontaneous macroscopic order).
The superradiant phase transition enables applications such as controllable non-reciprocal light-matter interaction \cite{nonrep}, quantum metrology \cite{quantmetro}, and chiral edge flow generation \cite{quantmetro}. Its study has recently expanded to diverse platforms including Rydberg atoms \cite{dipole}, Bose–Einstein condensates \cite{baumann2010dicke}, Fermi gases \cite{zhaihui2014,wuhaibin}, and magnons \cite{magnons}.
%The engineering of the superradiant phase transition has been demonstrated to have various applications, including the achievement of controllable non-reciprocal light-matter interaction \cite{nonrep}, quantum metrology  \cite{quantmetro}, and the generation of chiral edge flows \cite{chiral}. Recently, the study of superradiant phase transition has been extended to a variety of fields, including Rydberg atoms \cite{dipole}, Bose–Einstein condensates \cite{baumann2010dicke}, Fermi gases \cite{zhaihui2014,wuhaibin} and magnons \cite{magnons}.

The surface acoustic wave phonon propagates at a speed that is 5 orders of magnitude slower than the speed of light, which offers unique advantages for quantum state manipulation \cite{quantumphononics} in solid-state system. Recent experimental success in coupling a surface acoustic wave to a superconducting qubit \cite{quantuacoustic} has led to a new paradigm for on-chip quantum information processing \cite{quantuminfor}, precision measurement and quantum sensing \cite{senor,quantumphononics}. SAW couples to solid-state emitters (such as Nitrogen-Vacancy (NV) centers) via strain or piezoelectric effects and can mediate long-range interactions  \cite{wanghailin,saw2}, making them ideal for exploring collective effect of many-body physics \cite{brandes2005coherent} and phonon radiation \cite{phonlasr1,phonlasr2,phonlasr3,phonlasr4,phonlasr5,phonlasr6}. The cooperative coupling of multiple quantum emitters to a SAW to maintain coherence is a prerequisite for the establishment of a quantum network on a chip \cite{quantunet}. Therefore, there is an urgent requirement for an efficient scheme for the implementation of SAW superradiant phase for integrated chip-scale quantum state manipulation.

In the Dicke model \cite{PhysRev.93.99}, the threshold of atom-cavity coupling strength $g_c$ for the occurrence of superradiance is in the strong coupling regime \cite{RN4,PhysRevA.7.831}. However, achieving strong coupling remains challenging. Reducing critical coupling strength in order to achieve superradiance in a weak coupling regime is an ongoing research priority \cite{squeezesup,dipole}. Furthermore, in most cases, emitter-radiation coupling is an inherent property of the system. Once the system has been determined, it is typically challenging to adjust the coupling strength using external control methods. In solid-state systems, the strain-induced coupling between phonons (SAW) and quantum defects (NV centers) \cite{RN4, macquarrie2013mechanical} is typically faint. The cooperative coupling of multiple quantum defects in solid-state system with SAW to achieve superradiance with an external tunable approach represents a significant challenge.

In this paper, we propose an optically tunable scheme to induce SAW phonon superradiance in an ensemble of NV centers in diamond under weak coupling. Laser driving of the NV center's electronic transitions, assisted by phonon absorption or emission, creates an effective spin-phonon-photon coupling. By tuning the laser intensity, a phonon superradiant phase transition occurs in the weak coupling region. Moreover, the transition occurs instantaneously when the driving light exceeds a certain threshold, and remains effective for a limited number of particles. Our work provides a clear and experimentally feasible method for observing this transition with an optically tunable approach, paving the way for chip-scale quantum networks.

%In this paper, we propose a driven-dissipative scheme to induce a SAW phonon superradiance in an ensemble of NV centers embedded in diamond operating within the weak coupling regime. The use of a laser to drive the NV center's electronic levels transition, assisted by phonon absorption or emission, enables the realization of an effective spin-phonon-photon coupling. Tuning the driving laser intensity results in a SAW phonon superradiant phase transition occurring at a weak coupling region. Furthermore, to investigate the dynamics of the phase transition, we show that this transition occurs instantaneously when the driving light is activated beyond a certain threshold. For a limited number of particles, this optical switched phonon superradiant phase transition remains effective. Our work proposes a clear and experimentally feasible scheme for observing the SAW phonon superradiant transition in the weak coupling region with an optical tunable method, with the potential to enable the implementation of a quantum network on a chip scale.

The remainder of this paper is organized as follows. In Sec. \ref{sec:model}, we present the model and derive the effective Hamiltonian. In Sec. \ref{semi}, we employ the mean-field method to obtain the steady-state phonon amplitude and the threshold coupling for the superradiant phase transition. In Sec. \ref{sec:switch}, we demonstrate the dynamics of the optically tunable superradiant phase transition. In Sec. \ref{sec:finiteN}, we investigate this transition in finite-size particle systems. Finally, we give a summary in Sec. \ref{sum}.

%The paper is organized as follows. In section \ref{sec:model}, the model is presented and the effective Hamiltonian is derived. In section \ref{semi}, the equations of motion are solved through the utilization of the mean-field method, thus yielding the steady-state phonon field amplitude and threshold of coupling strength of the superradiant phase transition. In section \ref{sec:switch}, the dynamics of the optical tunable phonon superradiant phase transition is demonstrated. In section \ref{sec:finiteN}, we explore the effective optical tuning of the SAW phonon superradiant phase transition in a limited number of particle systems. Finally, in section \ref{sum}, we give a summary.

\section{MODEL AND HAMILTONIAN}\label{sec:model}

\begin{figure}[t!]
    \includegraphics[width=\columnwidth]{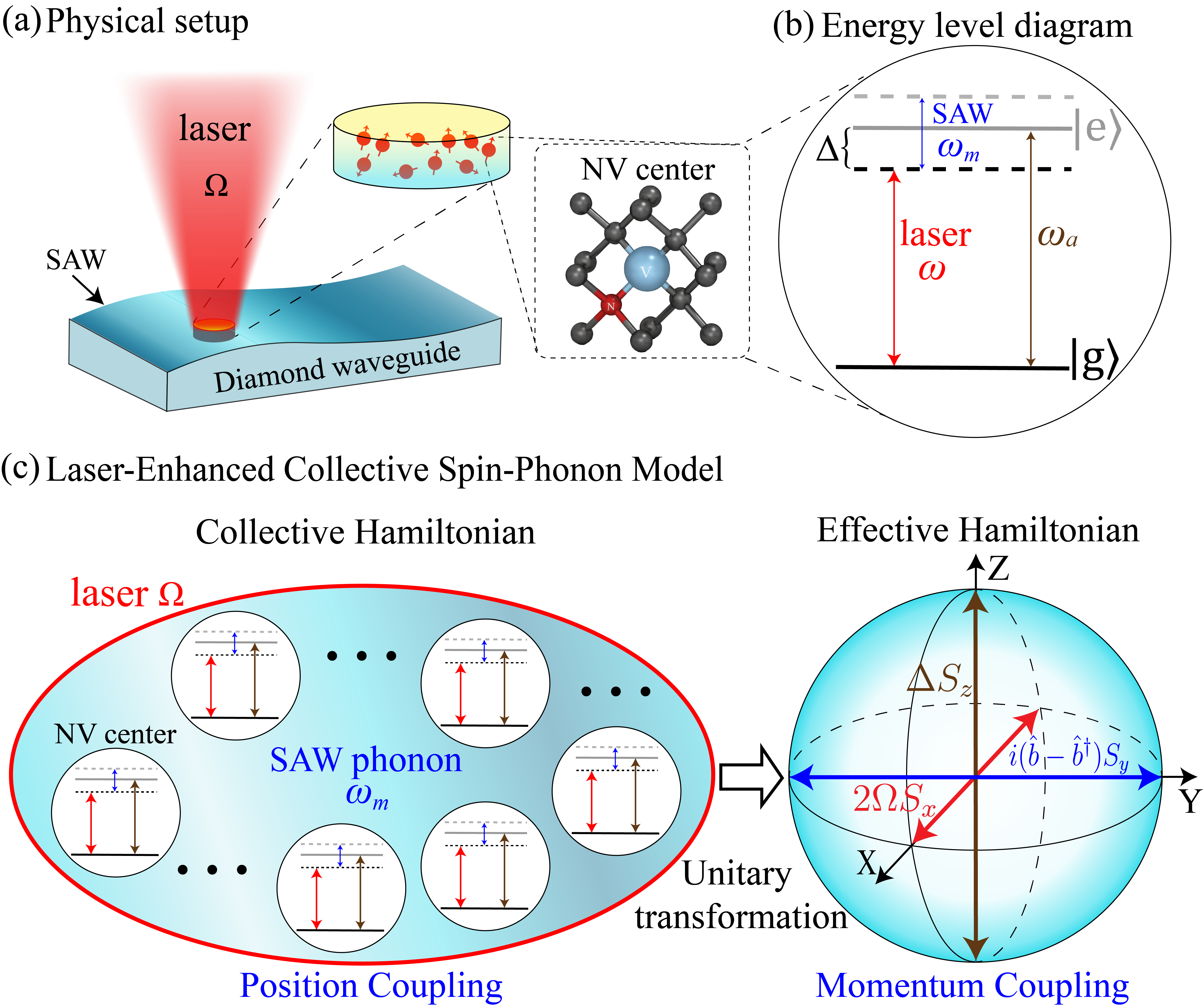} 
    \caption{
    (a)~Schematic setup: an ensemble of NV centers in a diamond waveguide is coupled to a single SAW mode and driven by a classical laser with coupling strength $\Omega$. 
    (b)~Single-NV energy diagram. The laser drives the $|g\rangle \leftrightarrow |e\rangle$ transition with detuning $\Delta=\omega_a-\omega$, while the SAW strain modulates the excited-state energy.[Eq.~(\ref{eq:H_single})]
    (c)~Collective and effective spin--phonon models. The original collective Hamiltonian [Eq.~(\ref{eq:H_collective})] describes uniformly driven NV centers coupled to a single SAW mode. The effective Hamiltonian [Eq.~(\ref{eq:H_eff})] contains the three terms $\Delta S_z$, $2\Omega S_x$, and $i(\hat{b}-\hat{b}^\dagger)S_y$, with $G=\lambda\Omega/\omega_m$.
    }
    \label{fig:model_schematic}
\end{figure}

\begin{comment}
\begin{figure}[t!]
    \includegraphics[width=\columnwidth]{fig1_phonon superradience model_v11.png} 
    \caption{
    (a)~Physical setup: an ensemble of Nitrogen-Vacancy (NV) centers embedded in a diamond waveguide are collectively coupled to a surface acoustic wave (SAW) and driven by a classical laser field with optical drive strength $\Omega$. The inset shows the atomic structure of an NV center.
    %
    (b)~Energy level diagram of a single NV center illustrating the microscopic ingredients entering Eq.~(\ref{eq:H_single}).
    The laser with frequency $\omega$ drives the transition from the ground state $|g\rangle$ to the excited state $|e\rangle$ with detuning $\Delta=\omega_a-\omega$, and the SAW phonon of frequency $\omega_m$ modulates the excited-state level via strain coupling.
    %
    (c)~Laser-enhanced collective and effective spin-phonon model. The collective Hamiltonian [Eq.~(\ref{eq:H_collective})] describes the uniform coupling of $N$ NV centers to a single-mode SAW phonon, with all emitters driven by the same classical laser. After the state-dependent displacement and rotating-frame transformations, the effective Hamiltonian $H_{\rm eff}$ [Eq.~(\ref{eq:H_eff})] is represented in the collective spin space, where the three orthogonal axes correspond to the terms $\Delta S_z$, $2\Omega S_x$, and $i(\hat{b}-\hat{b}^\dagger)S_y$, respectively, with $G=\lambda\Omega/\omega_m$.
    }
    \label{fig:model_schematic}
\end{figure}
\end{comment}

The system (Fig.~\ref{fig:model_schematic}(a)) consists of $N$ identical NV centers in a 1D diamond waveguide. Each NV center is a two-level qubit with ground state $|g\rangle$, excited state  $|e\rangle$, and transition frequency $\omega_a$ [Fig.~\ref{fig:model_schematic}(b)].
%The whole system, as depicted schematically in Fig.~\ref{fig:model_schematic}(a), comprises $N$ identical NV centers embedded in a one-dimensional diamond waveguide. Each NV center is modeled as a two-level qubit with ground state $|g\rangle$ and excited state $|e\rangle$, separated by a transition frequency $\omega_a$ [Fig.~\ref{fig:model_schematic}(b)]. 
For a single NV center driven by a classical light (frequency $\omega$, coupling strength $\Omega$) and coupled to a SAW phonon field (coupling strength $\lambda$), the Hamiltonian is $H=H_0+H_{\rm s\text{-}ph}+H_{\rm drive}$, with \cite{PhysRevA.7.831}
\begin{align}
H_0 &= \omega_m \hat{b}^\dagger \hat{b}+\frac{\omega_a}{2}\sigma_z, \notag\\
H_{\rm s\text{-}ph} &= \frac{\lambda}{\sqrt{V}}(\hat{b}+\hat{b}^\dagger)|e\rangle\langle e|, \notag\\
H_{\rm drive} &= \Omega\left(e^{-i\omega t} \sigma_{+}+e^{i\omega t} \sigma_{-}\right),
\label{eq:H_single}
\end{align}
where $\sigma_z=|e\rangle\langle e|-|g\rangle\langle g|$, $\sigma_{+}=\sigma_{-}^\dagger=|e\rangle\langle g|$ and $V$ is the volume of phonon field. Here, $H_0$ is the free Hamiltonian of the SAW phonon and the NV center qubit, $H_{\rm s\text{-}ph}$ is the excited-state-dependent spin-phonon coupling, and $H_{\rm drive}$ describes the coherent optical drive between $|g\rangle$ and $|e\rangle$. For NV centers, the excited-state strain coupling is about five orders of magnitude stronger than that of the ground state \cite{saw1,RN4,wanghailin,PhysRevB.88.064105}; thus, the spin-phonon interaction can be well approximated by the strain-induced energy shift of the excited state alone.
%Here $H_0$ is the free Hamiltonian of the SAW phonon mode and the bare two-level NV transition, $H_{\rm s\text{-}ph}$ is the excited-state-dependent dispersive spin-phonon coupling, and $H_{\rm drive}$ represents the coherent optical drive between $|g\rangle$ and $|e\rangle$. For NV centers, the excited-state strain coupling is approximately five orders of magnitude stronger than the ground-state one \cite{saw1,RN4,wanghailin,PhysRevB.88.064105}; therefore, the spin-phonon interaction is well approximated by the strain-induced energy shift of the excited state alone.

Assuming homogeneous coupling and fixed NV center's density $\rho=N/V$, the collective Hamiltonian of the $N$-emitter ensemble, illustrated in the left part of Fig.~\ref{fig:model_schematic}(c), can be written as the collective Hamiltonian:
\begin{equation}
\begin{aligned}
H &= \omega_m \hat{b}^{\dagger}\hat{b}
+ \omega_a S_z
+ \frac{\lambda}{\sqrt{N}}
\left(\hat{b}+\hat{b}^{\dagger}\right)P_e \\
&\quad
+ \Omega\left(
e^{-i\omega t}S_+
+ e^{i\omega t}S_-
\right).
\end{aligned}
\label{eq:H_collective}
\end{equation}
where the collective operators are $S_\alpha=\frac{1}{2}\sum_i \sigma_\alpha^i$ ($\alpha=x,y,z$), $S_\pm=\sum_i \sigma_\pm^i$, and $P_e\equiv\sum_i |e_i\rangle\langle e_i|=N/2+S_z$. In contrast to the single-emitter Hamiltonian (\ref{eq:H_single}), the current Hamiltonian treats the collective NV centers as a single large spin coupled to the driving field and the SAW phonon.
%The first two terms describe the free Hamiltonian of the phonon mode and the collective spins, the third term describes the strain-induced coupling between the phonon displacement and the excited-state population, and the last term is the coherent laser drive. 
Unlike the standard Dicke model \cite{PhysRev.93.99}, the SAW couples to the excited-state operator $P_e$ rather than directly to $(\sigma_+^i+\sigma_-^i)$.

Using the state-dependent displacement transformation $U_d=\exp[-\lambda(\hat{b}^\dagger-\hat{b})P_e/(\sqrt{N}\omega_m)]$ and then the rotating-frame transformation $U_r(t)=\exp(-i\omega t P_e)$, and retaining terms to first order in $\lambda/(\sqrt{N}\omega_m)$, we obtain the effective Hamiltonian shown in the right part of Fig.~\ref{fig:model_schematic}(c)
\begin{equation}
H_{\rm eff} = \omega_m \hat{b}^\dagger \hat{b} + \Delta\, S_z + 2\Omega\, S_x + i\frac{2G}{\sqrt{N}}(\hat{b}-\hat{b}^\dagger) S_y,
\label{eq:H_eff}
\end{equation}
where $\Delta=\omega_a-\omega$ is the renormalized detuning and $G=\lambda\Omega/\omega_m$ is the laser-enhanced spin-phonon coupling strength. Introducing the phonon quadrature operators $\hat{X}_{\rm ph}=(\hat{b}+\hat{b}^\dagger)/\sqrt{2}$ and $\hat{P}_{\rm ph}=i(\hat{b}^\dagger-\hat{b})/\sqrt{2}$, the interaction term may be written as $(2\sqrt{2}G/\sqrt{N})\hat{P}_{\rm ph}S_y$, showing that the original position-like coupling $\hat{X}_{\rm ph}P_e$ is mapped to a momentum-like coupling (For a detailed derivation, see the Appendix \ref{app:model}).

Equation~(\ref{eq:H_eff}) contains two transition channels: the laser-induced phonon-assisted channel $i(2G/\sqrt{N})(\hat{b}-\hat{b}^\dagger)S_y$ and the direct optical driving channel $2\Omega S_x$. Since $G=\lambda\Omega/\omega_m\propto\Omega$, the driving laser directly controls the effective spin-phonon coupling and can push the system into the superradiant phase even when the bare spin-phonon coupling $\lambda$ is weak.

\section{Optically Tunable Superradiant Phase Transition}\label{semi}

The system under discussion allows energy to leak into the environment. In addition, the levels of the NV centers are driven by light (Fig. \ref{fig:model_schematic} (a)), thus creating a driven-dissipative system. The evolution of this system is governed by the master equation $\dot{\rho} = -i[H_{eff}, \rho] + \mathcal{L}_{ph}(\rho) + \mathcal{L}_{NV}(\rho)$, where the Lindbladian terms $\mathcal{L}_{ph}(\rho) = \kappa(2\hat{b}\rho \hat{b}^\dagger-\hat{b}^\dagger \hat{b} \rho - \rho \hat{b}^\dagger \hat{b})$ and $\mathcal{L}_{NV}(\rho) = \gamma(2S_{-}\rho S_{+}-S_{+} S_{-} \rho - \rho S_{+} S_{-} )$ respectively account for the dissipations of the phonons and qubits. In our system, we assume that the decay rate of the qubits $\gamma$ are much smaller than the phonon field decay rate $\kappa$. In the subsequent analytical derivations, the value of $\gamma$ is set to $0$ in order to ensure mathematical simplicity. Furthermore, the conservation of spin is a process of considerable significance.

\begin{figure}[t!]
    \centering
    \includegraphics[width=1.0\columnwidth]{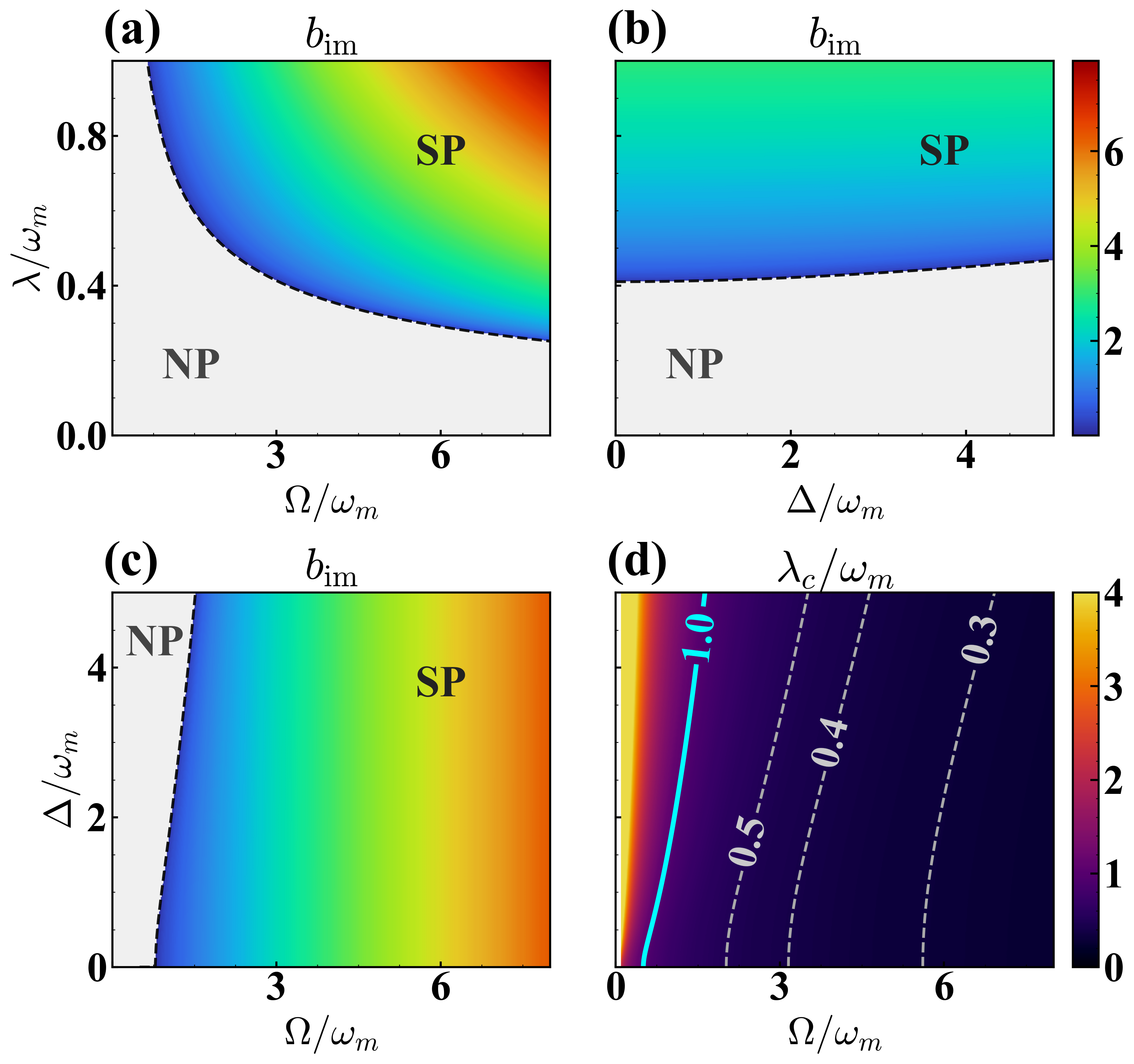}
    \caption{Mean-field phase diagrams of SAW phonon superradiant transition. (a)--(c) Imaginary phonon amplitude $b_{\rm im}$ (color scale) versus (a) $\Omega/\omega_m$ (light-qubit coupling strength) and $\lambda/\omega_m$ (phonon-qubit coupling strength) at $\Delta/\omega_m=1.0$ (light-qubit detuning), (b) $\Delta/\omega_m$ and $\lambda/\omega_m$ at $\Omega/\omega_m=3.0$, (c) $\Omega/\omega_m$ and $\Delta/\omega_m$ at $\lambda/\omega_m=0.6$. Gray: normal phase (NP, $b_0=0$); colored: superradiant phase (SP, $b_0\neq0$). Black dashed curves: critical coupling strength $\lambda_c$ [Eq.~(\ref{eq:critical_coupling})]. (d) $\lambda_c$ in the $(\Omega,\Delta)$ plane. Cyan: $\lambda_c=1.0\,\omega_m$; gray dashed: $\lambda_c=0.3,0.4,0.5\,\omega_m$. Other parameter: $\kappa=0.1\,\omega_m$.} 
       
       % Mean-field phase diagrams of the SAW phonon superradiance.(a)--(c)~Imaginary part of the phonon-field amplitude $b_{\rm im}$ (color scale) in different parameter planes; gray regions denote the normal phase (NP, $b_0=0$) and colored regions the superradiant phase (SP, $b_0\neq 0$). Black dashed curves mark the analytical phase boundary $\lambda_c$ [Eq.~(\ref{eq:critical_coupling})]. Phonon-field amplitudes as functions of (a) $\Omega$ and $\lambda$ at fixed $\Delta/\omega_m=1.0$; %Crucially, even in the weak-coupling regime $\lambda<\omega_m$, the SP becomes accessible by increasing the laser drive $\Omega$---the central feature of our optically switched scheme.(b) $\Delta$ and $\lambda$ at fixed $\Omega/\omega_m=3.0$;(c) $\Omega$ and $\Delta$ at fixed $\lambda/\omega_m=0.6$ are shown.(d)~Critical coupling $\lambda_c$ [Eq.~(\ref{eq:critical_coupling})] over the $(\Omega,\Delta)$ plane. The cyan solid line marks $\lambda_c=1.0\,\omega_m$; gray dashed contours indicate $\lambda_c=0.3$, $0.4$, and $0.5$ $\omega_m$. Darker colors correspond to smaller $\lambda_c$, i.e., easier access to the SP.Other parameter: $\kappa=0.1\,\omega_m$.
    \label{fig:phase_diagrams}
\end{figure}

By applying the mean-field approximation (e.g., $\langle S_x \hat{b} \rangle \approx \langle S_x \rangle \langle \hat{b} \rangle$), the Heisenberg--Langevin equations for the operator mean values are given by
\begin{equation}
\left\{
\begin{aligned}
    \frac{d\langle \hat{b}\rangle}{dt} &= -i\omega_m \langle \hat{b}\rangle - \frac{2G}{\sqrt{N}}\langle S_y \rangle - \kappa\langle \hat{b}\rangle ,\\
    \frac{d\langle S_x\rangle}{dt} &= -\Delta\langle S_y \rangle +i\frac{2G}{\sqrt{N}} (\langle\hat{b}\rangle -\langle \hat{b}^\dagger \rangle)\, \langle S_z \rangle ,\\
    \frac{d\langle S_y\rangle}{dt} &= \Delta\langle S_x \rangle - 2\Omega\langle S_z \rangle ,\\
    \frac{d\langle S_z\rangle}{dt} &= 2\Omega\langle S_y \rangle -i \frac{2G}{\sqrt{N}} (\langle\hat{b}\rangle -\langle \hat{b}^\dagger \rangle)\, \langle S_x \rangle . \label{eq:eom}
\end{aligned}
\right. 
\end{equation}
In the steady state, it is convenient to introduce the normalized collective-spin variables $X=\langle S_x\rangle/N$, $Y=\langle S_y\rangle/N$, and $Z=\langle S_z\rangle/N$. We also define the rescaled phonon field amplitude $b_0=\langle\hat{b}\rangle/\sqrt{N}=b_{\rm re}+i\,b_{\rm im}$, where $b_{\rm re}=\mathrm{Re}\,b_0$ and $b_{\rm im}=\mathrm{Im}\,b_0$ are the real and imaginary parts of the normalized phonon amplitude.% The corresponding mean-field quadratures are $Q=\sqrt{2}\,b_{\rm re}$ and $P=\sqrt{2}\,b_{\rm im}$.
Solving the steady-state equations in terms of these variables yields the phonon order parameter.

The superradiant phase is characterized by a non-trivial solution $b_0\neq 0$, and this condition defines the phase boundary. Solving for $Z$ gives the critical value required for the onset of superradiance, $Z_c = -(\kappa^2+\omega_m^2)\Delta/(8G^2\omega_m)$. Combined with the spin conservation law $X^2+Y^2+Z^2=1/4$, the steady-state phonon amplitude---which serves as the order parameter---is
\begin{equation}
    \begin{cases}
    \displaystyle b_{\rm re} = - \frac{\kappa}{\omega_m}\, b_{\rm im}, \\[6pt]
    \displaystyle b_{\rm im} = \pm \frac{\Delta}{4G} \sqrt{\frac{1}{4Z_c^2} - 1 - \frac{4\Omega^2}{\Delta^2}}.
    \end{cases}
    \label{eq:order_parameter}
\end{equation}
Equations~\eqref{eq:order_parameter} determine both the phase boundary and the order parameter of the superradiant phase transition.

Figure~\ref{fig:phase_diagrams} highlights the optical tunability of the transition. Since the effective spin-phonon coupling satisfies $G=\lambda\Omega/\omega_m$ [Eq.~\eqref{eq:H_eff}], increasing the optical drive coupling strength $\Omega$ directly enhances the effective interaction. Consequently, Fig.~\ref{fig:phase_diagrams}(a) shows that stronger optical driving reduces the bare phonon-NV coupling $\lambda$ required for entering the superradiant phase, allowing superradiance even in the weak-coupling regime $\lambda/\omega_m<1$. At $\Omega=0$, the effective coupling vanishes and the phonon mode is decoupled from the NV ensemble, so the system remains in the normal phase. Figure~\ref{fig:phase_diagrams}(b) shows that the phase boundary depends only weakly on $\Delta$, while Fig.~\ref{fig:phase_diagrams}(c) shows that, for fixed $\lambda$, a threshold optical drive is sufficient to trigger the transition.

The critical bare coupling is (see Appendix~\ref{app:lambda} for details)
\begin{equation}
    \lambda_c = \frac{\sqrt{(\kappa^2 + \omega_m^2)\omega_m}}{2\Omega} \sqrt[4]{4\Omega^2 + \Delta^2}.
    \label{eq:critical_coupling}
\end{equation}
For $\Omega\gg\omega_m$, it scales as $\lambda_c\sim1/\sqrt{2\Omega}$, confirming that stronger optical driving lowers the threshold for SAW phonon superradiant phase transition (Fig.~\ref{fig:phase_diagrams}(d)).

%\section{Optically switched Phonon Superradiant Switch}
\section{Optically Tunable Phonon Superradiant Dynamics}
\label{sec:switch}

\begin{figure}[t!]
    \centering
    \includegraphics[width=\columnwidth]{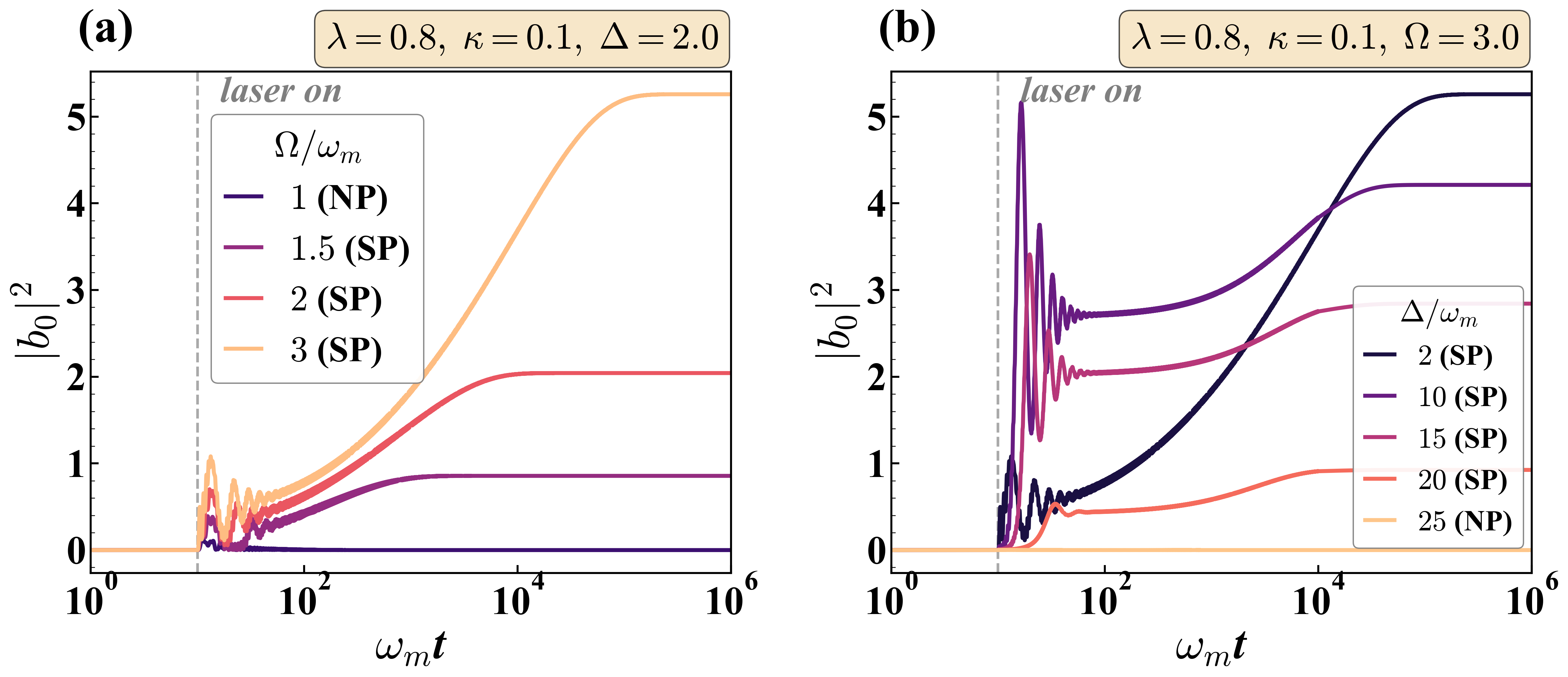}
    \caption{ Dynamics of an optically controlled phonon superradiant switch. Time evolution of rescaled phonon number $|b_0|^2$ after laser turn-on at $t_{\rm on}=50/\omega_m$ (dashed vertical line) for: (a) different $\Omega$ at fixed $\lambda/\omega_m=0.8$, $\Delta/\omega_m=2.0$, $\kappa/\omega_m=0.1$; (b) different $\Delta$ at fixed $\lambda/\omega_m=0.8$, $\Omega/\omega_m=3.0$, $\kappa/\omega_m=0.1$.
       % Optically controlled phonon superradiant switch. The time evolution of phonon fields ($|b_0|^2$) following the initiation of the laser at a time $t_{\rm on}=50/\omega_m$ (dashed vertical line) is shown in two scenarios:
        %(a)~varying light-qubit strength $\Omega$ at fixed 
        %$\lambda/\omega_m=0.8$, $\Delta/\omega_m=2.0$, $\kappa/\omega_m=0.1$;
       % and (b)~varying detuning $\Delta$ at fixed 
       % $\lambda/\omega_m=0.8$, $\Omega/\omega_m=3.0$, $\kappa/\omega_m=0.1$.
    }
    \label{fig:switch}
\end{figure}

The optically tuned critical threshold of the SAW phonon superradiant phase transition provides a rapid phonon switch. When the laser is off ($\Omega=0$), the effective coupling $G=\lambda\Omega/\omega_m$ vanishes and the system resides in the normal phase with $b_0=0$ regardless of $\lambda$. Switching the laser on with parameters satisfying $\lambda>\lambda_c$ drives the system into the superradiant phase with rescaled phonon number $|b_0|^2>0$. Despite the underlying continuous second-order phase transition (see  Appendix~\ref{app:fluctuations}), the laser on/off operation thus functions as a binary switch for the SAW phonon field.

We demonstrate the optically tunable phonon superradiance switch-on dynamics by numerically integrating the mean-field equations [Eq.~(\ref{eq:eom})] with the laser abruptly turned on at $t_{\rm on}=50/\omega_m $. Before $t_{\rm on}$, the system is in the normal phase (NP, $b_0=0$, $Z=-1/2$); after $t_{\rm on}$, % a small disturbance seeds the instability. 
the system initiates a transition to a superradiant phase (SP).
Figure~\ref{fig:switch}(a) shows the time evolution of $|b_0|^2$ for different coupling strengths $\Omega$ at fixed $\lambda/\omega_m=0.8$, $\Delta/\omega_m=2.0$, $\kappa/\omega_m=0.1$. For $\Omega/\omega_m\leq 1$ ($\lambda<\lambda_c$), the system remains in the normal phase (NP). For $\Omega/\omega_m\geq 2$ ($\lambda>\lambda_c$), it transitions to the superradiant phase (SP) over time. Increasing $\Omega$ ($\propto$ driving light intensity) enhances the relaxation rate and steady-state amplitude, consistent with $G\propto\Omega$, confirming laser intensity as the control knob. Figure~\ref{fig:switch}(b) displays the time dynamics for different $\Delta$ at fixed $\Omega/\omega_m=3.0$. All curves with $\Delta/\omega_m\leq 20$ reach the SP at steady state, while $\Delta/\omega_m=25$ remains in the NP. The steady-state amplitude decreases with $\Delta$, reflecting the reduced order parameter [Eq.~\ref{eq:order_parameter}]. Thus, driving near resonance is optimal for implementing the phonon switch.
%Figure~\ref{fig:switch}(a) shows $|b_0|^2$ versus time for different drive strengths $\Omega$ at fixed $\lambda/\omega_m=0.8$, $\Delta/\omega_m=2.0$, $\kappa/\omega_m=0.1$. For $\Omega/\omega_m\leq 1$ (where $\lambda<\lambda_c$), the system remains in the NP throughout the entire time range. For $\Omega/\omega_m\geq 2$ (where $\lambda>\lambda_c$), the system enters the SP as the time increases. Moreover, an increase in $\Omega$ results in larger growth rate and a higher steady-state amplitude, which is consistent with the enhanced effective coupling $G\propto\Omega$. This confirms that the laser intensity serves as the control knob for the phonon superradiant switch. Figure~\ref{fig:switch}(b) displays the time dynamics of the phonon field for different detuning $\Delta$ at fixed $\Omega/\omega_m=3.0$. All curves with $\Delta/\omega_m\leq 20$ lie within the SP in steady state, while $\Delta/\omega_m=25$ falls in the NP. The steady-state amplitude decreases with increasing $\Delta$, reflecting the reduced order parameter [Eq.~(\ref{eq:order_parameter})]. Consequently, it can be deduced that the driving light-spin resonance constitutes the optimal solution for the implementation of the phonon switch.

%\section{Few-N Effects on the Superradiant Crossover}
\section{Optically Tunable Finite-N Superradiance}
\label{sec:finiteN}

\begin{figure}[!t]
\centering
\includegraphics[width=\columnwidth]{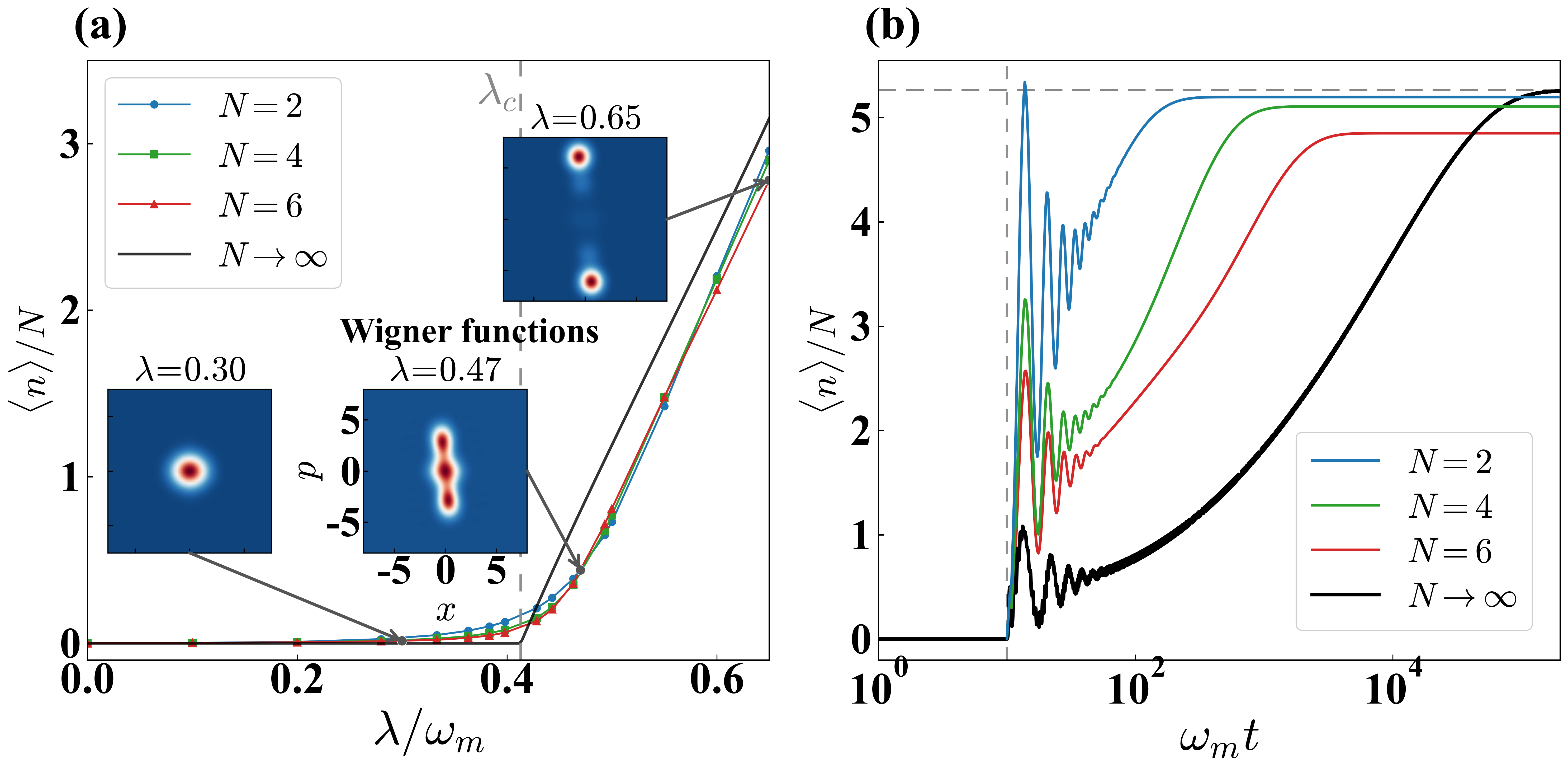}
\caption{Finite-$N$ effects on the superradiant crossover and switch dynamics.
(a)~Normalized phonon number $\langle n\rangle/N$ versus coupling $\lambda$ for $N=2,4,6$ and the mean-field limit. The dotted line marks $\lambda_c$. Insets show representative steady-state Wigner functions for $N=6$ in the NP, near the phase boundary, and deep in the SP.
(b)~Switch-on dynamics of $\langle n\rangle/N$ for $N=2,4,6$ and the mean-field solution. The laser is turned on at $t_{\rm on}=50/\omega_m$, and $\langle n\rangle/N\to |b_0|^2$ for $N\to\infty$.
Parameters: (a)~$\Delta/\omega_m=1$, $\Omega/\omega_m=3$, $\kappa=0.1\,\omega_m$;
(b)~$\lambda/\omega_m=0.8$, $\Delta/\omega_m=2$, $\Omega/\omega_m=3$, $\kappa=0.1\,\omega_m$.}
\label{fig:finiteN}
\end{figure}

\begin{comment}
\begin{figure}[!t]
\centering
\includegraphics[width=\columnwidth]{fig4_main_updated.png}
\caption{Finite-$N$ effects on the superradiant  crossover and switch dynamics.
(a)~Normalized phonon number $\langle n\rangle/N$ versus spin-phonon coupling $\lambda$ for $N=2,4,6$ (colored curves) and the mean-field limit $N\to\infty$ (solid black). The vertical dotted line marks $\lambda_c$. Insets: Wigner functions $W(\alpha)$ of the steady-state phonon field, with axes $\mathrm{Re}\,\alpha$ (horizontal) and $\mathrm{Im}\,\alpha$ (vertical), for the NP ($N=4$, $\lambda/\omega_m=0.30$, lower-left), near the phase boundary ($N=4$, $\lambda/\omega_m=0.50$, lower-middle), and deep in the SP ($N=4$, $\lambda/\omega_m=0.65$, upper-right).
(b)~Phonon switch-on dynamics: $\langle n\rangle/N$ versus time $t$ for $N=2,4,6$ and the mean-field solution (black). The laser is turned on at $ t_{\rm on}=50/\omega_m$. Note that for $N\to\infty$, $\langle n\rangle/N\to|b_0|^2$.
Parameters: (a)~$\Delta/\omega_m=1$, $\Omega/\omega_m=3$, $\kappa=0.1\,\omega_m$;
(b)~$\lambda/\omega_m=0.8$, $\Delta/\omega_m=2$, $\Omega/\omega_m=3$, $\kappa=0.1\,\omega_m$.}
\label{fig:finiteN}
\end{figure}
\end{comment}

The preceding analysis is based on the thermodynamic limit ($N,V\to\infty$ with $N/V$ constant), where the mean-field approximation becomes exact and the system undergoes a genuine second-order superradiant phase transition at $\lambda=\lambda_c$ (see Appendix~\ref{app:fluctuations} for details).
% A sharp phase transition, however, requires the non-analyticity of the order parameter that can develop only in the limit of infinitely many degrees of freedom. For a finite ensemble the steady-state density matrix is finite-dimensional and all observables remain analytic functions of $\lambda$, so no true singularity---and hence no genuine symmetry breaking---can occur: the $\mathbb{Z}_2$ symmetry ($\hat b\to-\hat b$, $S_y\to-S_y$) is never spontaneously broken, and the normal-phase vacuum survives with finite weight $p_0$ alongside the superradiant configurations (Appendix~\ref{app:finiteN}).The sharp transition is therefore replaced by a smooth crossover. 
For a realistic system, the number of NV centers is finite (potentially as few as $N\sim 2$--$10$). Whether the finite-size effect from such a finite number affects the optically tunable phonon superradiant transition remains to be determined. Using exact numerical solutions of the master equation, we show the phonon superradiant switch functionality persist in small systems, albeit with quantitative modifications. Moreover, for finite $N$, the quantum phase transition is smoothed by a phase crossover, and it sharpens monotonically toward the mean-field transition as $N$ increases.
%This matters for any realistic SAW-NV implementation, where only a limited number $N$ of NV centers is involved, potentially as few as $N\sim 2$--$10$. It is therefore essential to examine whether the optically tuned superradiant crossover and the phonon switch remain operational at finite $N$. Using exact numerical solutions of the master equation, we show that both the crossover and the switch functionality persist in small systems, albeit with quantitative modifications; the crossover sharpens monotonically toward the mean-field transition as $N$ grows.

Figure~\ref{fig:finiteN}(a) plots the normalized phonon number $\langle n\rangle/N$ versus coupling $\lambda$ for $N=2,4,6$, along with the mean-field solution (solid black curve) as $N\to\infty$; the transition sharpens monotonically with $N$, approaching the mean-field limit, confirming the asymptotic exactness of the mean-field treatment in the thermodynamic limit. The insets show the steady-state Wigner functions $W(\alpha)$: a single Gaussian at the origin in the NP (vacuum behavior), two dominant superradiant peaks in the SP, and near the phase boundary, a three-peak structure (central vacuum peak coexisting with two displaced peaks) reflecting finite-$N$ coexistence of NP and SP configurations. This three-peak Wigner function is a uniquely quantum signature absent from mean-field theory, and its observation via quantum state tomography~\cite{RN2} would directly evidence quantum-level coexistence of vacuum and symmetry-broken superradiant states.

Figure~\ref{fig:finiteN}(b) compares the switch-on dynamics of $\langle n\rangle/N$ for $N=2,4,6$ with the mean-field solution. For small $N$ ($N=2$), the phonon field amplitude exhibits more pronounced oscillations due to the larger quantum fluctuations inherent in smaller systems. As time increases, the oscillating phonon fields approach a steady-state value more rapidly for smaller values of $N$. As $B$ increases, the steady-state phonon field amplitude converges to a fixed value near the mean-field result. Importantly, the switch-on functionality is preserved for all $N$, as evidenced by the phonon field remaining zero before the laser is switched on and growing to a finite steady state value after the laser is turned on. This confirms that the optically controlled phonon switch is robust against finite-size effects.

In recent experiments, coherent SAW-NV coupling has been demonstrated at $\omega_m/2\pi=818$~MHz~\cite{saw1}, with a single-emitter coupling rate $g_0/2\pi\sim 10$~MHz~\cite{saw1,quantunet}. This rate originates from the large excited-state orbital strain susceptibility $d_{\rm es}\approx 1$~PHz/strain~\cite{ovartchaiyapong2014,macquarrie2017}, which exceeds the ground-state spin--strain coupling by roughly five orders of magnitude~\cite{PhysRevLett.113.020503}. SAW cavities on diamond with quality factors $Q=10^3$--$10^5$ have been reported~\cite{quantuacoustic,saw2}, corresponding to phonon decay rates $\kappa/2\pi\sim 10$~kHz--$1$~MHz. The collective spin-phonon coupling strength scales as $\lambda=g_0\sqrt{N}$, so an ensemble of $N\sim 25$ NV centers yields $\lambda/2\pi\sim 50$~MHz, while $N\sim 2500$ reaches $\lambda/2\pi\sim 500$~MHz --- well into the superradiant regime when assisted by a moderate laser drive. These parameters, combined with the finite-$N$ results of Fig.~\ref{fig:finiteN}, establish the optically controlled SAW phonon superradiant switch as a feasible building block for on-chip quantum acoustics and quantum communication.

\FloatBarrier
\section{Summary}\label{sum}
In summary, we have proposed a scheme for optically controllable generation of a SAW phonon superradiant phase transition in the weak-coupling region. Utilizing a laser to drive an ensemble of NV centers embedded in diamond enables optically driven electronic transition to assist SAW phonon emission, thereby generating a SAW phonon superradiance with a weak spin-phonon coupling strength. Furthermore, the quantum phase transition of a small system has been characterized, and it shows that the optically tuned superradiant transition remains valid in realistic experimental conditions.
Optical control of the phonon superradiant transition in diamond provides a fast quantum switch for controlling on-chip quantum devices. Our work paves the way for advancements in quantum acoustics and on-chip quantum communication and networking \cite{quantunetwork}, quantum metrology \cite{quantmetro, PhysRevA.93.022103}, as well as the engineering of non-classical phononic states \cite{quantumphononics}, by providing a highly effective path to harnessing collective spin-phonon interactions in solid-state quantum systems.
\FloatBarrier

\begin{acknowledgments}
This work is supported by the National Key R\&D Program of China (Grant No. 2022YFA1404400, No. 2023YFA1406900), the Natural Science Foundation of Shanghai (No. 23ZR1481200), the Program of Shanghai Academic Research Leader (No. 23XD1423800), the National Natural Science Foundation of China (No. 11935010), and the Opening Project of Shanghai Key Laboratory of Special Artificial Microstructure Materials and Technology.
\end{acknowledgments}
% ---- 补充材料 ----
\twocolumngrid
\appendix
\renewcommand{\thefigure}{A\arabic{figure}}
\setcounter{figure}{0} 

\section{Derivation of the effective Hamiltonian}
\label{app:model}

This appendix summarizes the intermediate steps connecting the microscopic Hamiltonian in Eq.~(\ref{eq:H_single}) to the effective collective Hamiltonian in Eq.~(\ref{eq:H_eff}). We keep only the terms needed in the main text.

For $N$ identical emitters with homogeneous coupling $g=\lambda/\sqrt{N}$, Eq.~(\ref{eq:H_single}) becomes
\begin{equation}
H=\omega_m \hat{b}^\dagger \hat{b}
+\omega_a S_z
+\frac{\lambda}{\sqrt{N}}(\hat{b}+\hat{b}^\dagger)P_e
+\Omega\left(e^{-i\omega t}S_+ + e^{i\omega t}S_-\right),
\label{eq:H_appendix_collective}
\end{equation}
where
\begin{equation}
P_e \equiv \sum_{i=1}^{N}|e_i\rangle\langle e_i| = \frac{N}{2}+S_z .
\label{eq:Pe_def}
\end{equation}

We first apply the state-dependent displacement transformation
\begin{equation}
U_d=\exp\!\left[-\eta(\hat b^\dagger-\hat b)P_e\right],
\qquad
\eta=\frac{\lambda}{\sqrt{N}\omega_m}.
\end{equation}
Using
\begin{equation}
U_d^\dagger \hat b U_d=\hat b-\eta P_e ,
\end{equation}
the phonon and strain-coupling terms are displaced as
\begin{align}
U_d^\dagger
\left[
\omega_m \hat b^\dagger \hat b
+\frac{\lambda}{\sqrt N}(\hat b+\hat b^\dagger)P_e
\right]
U_d
=
\omega_m \hat b^\dagger \hat b
-\frac{\lambda^2}{N\omega_m}P_e^2 .
\end{align}
Here \(P_e\) is the total excited-state population operator, not a projector for \(N>1\). Therefore \(P_e^2\neq P_e\) in general. Since the effective Hamiltonian used in the main text is retained only to first order in
\(\eta=\lambda/(\sqrt N\omega_m)\), the \(O(\lambda^2)\) term above is neglected consistently.

The optical drive transforms according to
\begin{align}
U_d^\dagger S_+ U_d
&=
S_+ e^{-\eta(\hat b^\dagger-\hat b)}, \\
U_d^\dagger S_- U_d
&=
e^{+\eta(\hat b^\dagger-\hat b)}S_- ,
\end{align}
where the sign convention is chosen consistently with the effective Hamiltonian in Eq.~(\ref{eq:H_eff}). The transformed Hamiltonian, up to \(O(\eta)\), is therefore
\begin{equation}
\begin{aligned}
U_d^\dagger H U_d \simeq
&\omega_m \hat b^\dagger \hat b
+\omega_a S_z\\
&+\Omega\left[
e^{-i\omega t}S_+e^{-\eta(\hat b^\dagger-\hat b)}
+
e^{i\omega t}e^{+\eta(\hat b^\dagger-\hat b)}S_-
\right].
\end{aligned}
\end{equation}

We then move to the rotating frame
\begin{equation}
U_r(t)=\exp(-i\omega t P_e).
\end{equation}
The rotated Hamiltonian \(H_r=U_r^\dagger(U_d^\dagger H U_d)U_r-iU_r^\dagger\dot U_r\) becomes
\begin{align}
H_r
&\simeq
\omega_m \hat b^\dagger \hat b
+\Delta S_z
+\Omega\left[
S_+e^{-\eta(\hat b^\dagger-\hat b)}
+
e^{+\eta(\hat b^\dagger-\hat b)}S_-
\right],
\end{align}
where, to the same order of approximation,
\begin{equation}
\Delta=\omega_a-\omega .
\end{equation}
Expanding the exponentials to first order,
\begin{equation}
e^{\pm\eta(\hat b^\dagger-\hat b)}
\simeq
1\pm\eta(\hat b^\dagger-\hat b),
\end{equation}
we obtain
\begin{align}
H_r
&\simeq
\omega_m \hat b^\dagger \hat b
+\Delta S_z
+\Omega(S_+ + S_-)
+\eta\Omega(\hat b^\dagger-\hat b)(S_- - S_+).
\end{align}
Using \(S_+ + S_- = 2S_x\) and \(S_- - S_+ = -2iS_y\), this gives
\begin{equation}
H_{\rm eff}
=
\omega_m \hat b^\dagger \hat b
+\Delta S_z
+2\Omega S_x
+i\frac{2G}{\sqrt N}
(\hat b-\hat b^\dagger)S_y ,
\label{heff}
\end{equation}
with
\begin{equation}
G=\frac{\lambda\Omega}{\omega_m}.
\end{equation}
This expression is identical to Eq.~(\ref{eq:H_eff}) in the main text.

Introducing the phonon quadratures
\begin{equation}
\hat X_{\rm ph}=\frac{\hat b+\hat b^\dagger}{\sqrt2},
\qquad
\hat P_{\rm ph}=\frac{i(\hat b^\dagger-\hat b)}{\sqrt2},
\end{equation}
the effective interaction (the last term in eq.(\ref{heff})) can be written as
\begin{equation}
\frac{2\sqrt2G}{\sqrt N}\hat P_{\rm ph}S_y .
\end{equation}
Thus, to leading order in the displacement parameter, the original position-like coupling to \(P_e\) is converted into an effective momentum-like spin-phonon coupling.

\section{Linear stability analysis and the three-stage switch-on dynamics}
\label{app:lambda}

This appendix derives the drift matrix $\mathbf A$ of the linearized mean-field equations, the critical coupling $\lambda_c$ from a rigorous stability analysis of the NP fixed point, and quantitatively characterizes the switch-on transient dynamics shown in Fig.~\ref{fig:anatomy} by tracking the trajectory in phase space.

\subsection{Normal-phase fixed point and drift matrix}

For completeness, we first write the mean-field equations of motion used in the following analysis. 
Under the mean-field approximation, the equations for the expectation values read
\begin{equation}
\left\{
\begin{aligned}
    \frac{d\langle \hat{b}\rangle}{dt} 
    &= -i\omega_m \langle \hat{b}\rangle
    - \frac{2G}{\sqrt{N}}\langle S_y \rangle
    - \kappa\langle \hat{b}\rangle ,\\
    \frac{d\langle S_x\rangle}{dt} 
    &= -\Delta\langle S_y \rangle
    +i\frac{2G}{\sqrt{N}} 
    \left(\langle\hat{b}\rangle -\langle \hat{b}^\dagger \rangle\right)
    \langle S_z \rangle ,\\
    \frac{d\langle S_y\rangle}{dt} 
    &= \Delta\langle S_x \rangle - 2\Omega\langle S_z \rangle ,\\
    \frac{d\langle S_z\rangle}{dt} 
    &= 2\Omega\langle S_y \rangle
    -i \frac{2G}{\sqrt{N}} 
    \left(\langle\hat{b}\rangle -\langle \hat{b}^\dagger \rangle\right)
    \langle S_x \rangle .
\end{aligned}
\right.
\label{eq:supp_eom}
\end{equation}

We use the normalized mean-field variables $b=\langle \hat b\rangle/\sqrt{N}=b_{\rm re}+i b_{\rm im}$ and $(X,Y,Z)=(\langle S_x\rangle,\langle S_y\rangle,\langle S_z\rangle)/N$, for which the spin length satisfies $X^2+Y^2+Z^2=1/4$.

Taking the real and imaginary parts of the phonon equation in Eq.~(\ref{eq:supp_eom}), the equations of motion take the explicit real form

\begin{equation}
\begin{aligned}
\dot b_{\rm re}&=-\kappa\,b_{\rm re}+\omega_m\,b_{\rm im}-2G\,Y, \\
\dot b_{\rm im}&=-\omega_m\,b_{\rm re}-\kappa\,b_{\rm im}, \\
\dot X&=-\Delta\,Y-4G\,b_{\rm im}\,Z, \\
\dot Y&=\Delta\,X-2\Omega\,Z, \\
\dot Z&=2\Omega\,Y+4G\,b_{\rm im}\,X,
\end{aligned}
\label{eq:eom_real}
\end{equation}
with $G=\lambda\Omega/\omega_m$. Here $Q=\sqrt 2\,b_{\rm re}$ and $P=\sqrt 2\,b_{\rm im}$ denote the mean-field counterparts of $\hat X_{\rm ph}$ and $\hat P_{\rm ph}$, respectively. Linearizing Eq.~(\ref{eq:eom_real}) around an arbitrary steady state $(b_{\rm re},b_{\rm im},X,Y,Z)$ by Taylor expansion to first order in $\delta\mathbf x=(\delta Q,\delta P,\delta X,\delta Y,\delta Z)^T$, the perturbations obey $\delta\dot{\mathbf x}=\mathbf A\,\delta\mathbf x$ with
\begin{equation}
\mathbf A=\begin{pmatrix}
-\kappa & \omega_m & 0 & -2\sqrt 2\,G & 0 \\
-\omega_m & -\kappa & 0 & 0 & 0 \\
0 & -2\sqrt 2\,G Z & 0 & -\Delta & -4G\,b_{\rm im} \\
0 & 0 & \Delta & 0 & -2\Omega \\
0 & 2\sqrt 2\,G X & 4G\,b_{\rm im} & 2\Omega & 0
\end{pmatrix},
\label{eq:drift_full}
\end{equation}
with $G=\lambda\Omega/\omega_m$.

\subsection{Critical coupling}

The normal-phase fixed point is obtained by setting $b_0=0$ and all time derivatives in Eq.~(\ref{eq:supp_eom}) to zero. This gives
\begin{equation}
\begin{aligned}
    Y_0& = 0,\\
    Z_0& = -\frac{\Delta}{2\sqrt{\Delta^2+4\Omega^2}},\\
    X_0& = \frac{2\Omega}{\Delta}\,Z_0 
    = -\frac{\Omega}{\sqrt{\Delta^2+4\Omega^2}}.
    \label{eq:fixed_point}
\end{aligned}
\end{equation}

Note that $Z_0 \neq -1/2$ whenever $\Omega \neq 0$: the laser drive provides a non-zero source term $-2\Omega Z$ in the equation for $\dot Y$ in Eq.~(\ref{eq:supp_eom}), which prevents $(X,Y,Z)=(0,0,-1/2)$ from being a steady state once the laser is on. The new fixed point is shifted along both $X$ and $Z$.

To determine the critical coupling $\lambda_c$, we specialize 
Eq.~(\ref{eq:drift_full}) to the NP fixed point by substituting 
$b_{\rm im}=0$, $X=X_0$, $Z=Z_0$. The $(3,5)$ and $(5,3)$ entries 
of Eq.~(\ref{eq:drift_full}) vanish, giving
\begin{equation}
\mathbf A_{\rm NP}
=\begin{pmatrix}\mathbf A_{\rm ph} & \mathbf A_{\rm sp\to ph}\\
                 \mathbf A_{\rm ph\to sp} & \mathbf A_{\rm sp}\end{pmatrix},
\label{eq:A_NP}
\end{equation}
where the four blocks are
\begin{align}
\mathbf A_{\rm ph}&=\begin{pmatrix}-\kappa & \omega_m\\ -\omega_m & -\kappa\end{pmatrix},
\quad
\mathbf A_{\rm sp}=\begin{pmatrix}0 & -\Delta & 0\\ \Delta & 0 & -2\Omega\\ 0 & 2\Omega & 0\end{pmatrix},\\[3pt]
\mathbf A_{\rm sp\to ph}&=\begin{pmatrix}0 & -2\sqrt 2\,G & 0\\ 0 & 0 & 0\end{pmatrix},
\quad
\mathbf A_{\rm ph\to sp}=\begin{pmatrix}0 & -2\sqrt 2\,G Z_0\\ 0 & 0\\ 0 & 2\sqrt 2\,G X_0\end{pmatrix}.
\label{eq:NP_blocks}
\end{align}
The diagonal block $\mathbf A_{\rm ph}$ has eigenvalues $-\kappa\pm i\omega_m$ (damped phonon oscillation at frequency $\omega_m$), and the antisymmetric $\mathbf A_{\rm sp}$ has eigenvalues $0,\pm i\omega_{\rm sp}$ with $\omega_{\rm sp}\equiv\sqrt{\Delta^2+4\Omega^2}$ (the zero eigenvalue reflects the conservation $X^2+Y^2+Z^2=1/4$). 
The NP becomes linearly unstable when one of the eigenvalues of 
$\mathbf A_{\rm NP}$ crosses zero. Direct expansion of 
$\det(\mathbf A_{\rm NP}-\mu\,\mathbf I)$ using Eq.~(\ref{eq:A_NP}) gives
\begin{equation}
\begin{aligned}
\det(\mathbf A_{\rm NP}-\mu\,\mathbf I)=
&-\mu\Bigl[\mu^4+2\kappa\mu^3+(\omega_{\rm sp}^2+\kappa^2+\omega_m^2)\mu^2\\
&+2\kappa\,\omega_{\rm sp}^2\,\mu-4\omega_m\,\omega_{\rm sp}\,(G^2-G_c^2)\Bigr],
\label{eq:char_poly}
\end{aligned}
\end{equation}
with
\begin{equation}
G_c^2\equiv\frac{\omega_{\rm sp}\,(\kappa^2+\omega_m^2)}{4\omega_m}.
\label{eq:Gc_def}
\end{equation}
The overall factor of $\mu$ in Eq.~(\ref{eq:char_poly}) is the trivial 
zero eigenvalue inherited from $\mathbf A_{\rm sp}$ and reflects the 
spin conservation $X^2+Y^2+Z^2=1/4$. The remaining four eigenvalues are 
roots of the quartic in the bracket. Setting $\mu=0$ in the bracket 
gives the threshold condition
\begin{equation}
G^2=G_c^2,
\end{equation}
i.e., the appearance of a second zero eigenvalue. Substituting 
$G=\lambda\Omega/\omega_m$ yields the critical coupling
\begin{equation}
\lambda_c=\frac{\sqrt{(\kappa^2+\omega_m^2)\,\omega_m}}{2\Omega}\,
(4\Omega^2+\Delta^2)^{1/4},
\label{eq:lambda_c_app}
\end{equation}
in agreement with Eq.~(\ref{eq:critical_coupling}) of the main text 
obtained from the order-parameter equation. For $G>G_c$, the bracket 
has a positive real root $\mu=\gamma>0$, and the NP fixed point becomes 
exponentially unstable along the corresponding eigenmode.

\subsection{Switch-on dynamics: numerical characterization}
\label{subsec:switch_dynamics}
\begin{figure*}[t!]
\centering
\includegraphics[width=\textwidth]{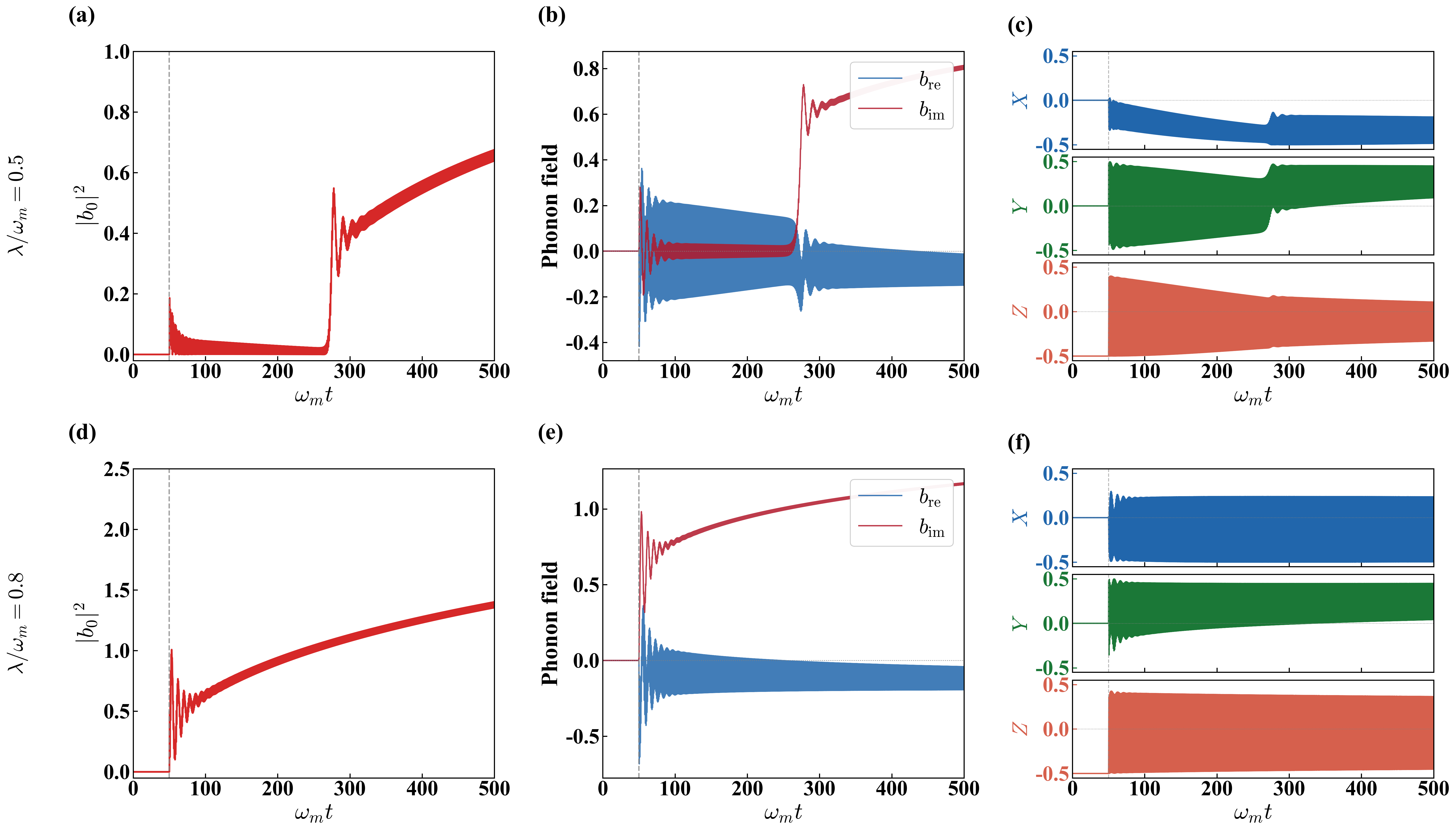}
\caption{Anatomy of the switch-on dynamics for $\lambda/\omega_m=0.5$ (top row) and $\lambda/\omega_m=0.8$ (bottom row). (a),(d)~$|b_0|^2$; Roman numerals in (a) mark the three stages discussed in the text. (b),(e)~Phonon quadratures $b_{\rm re}$ (blue), $b_{\rm im}$ (red). (c),(f)~Spin components $X$ (blue), $Y$ (green), $Z$ (red). Dashed line: $t_{\rm on} =50/\omega_m$. Parameters: $\Omega/\omega_m=3$, $\Delta/\omega_m=2$, $\kappa/\omega_m=0.1$.}
\label{fig:anatomy}
\end{figure*}

The dynamics of the spin-phonon interaction system can be obtained by solving the master equation. The temporal dynamics of the phonon field and collective spin are illustrated in Figure \ref{fig:anatomy}. For $\lambda/\omega_m=0.5$, the phonon field emerges from the vacuum state following the activation of the driving light at $t_{on} =50/\omega_m$. However, decoherence induced by environmental noise results in the annihilation process of the phonon field, as evidenced by the oscillatory damping of the phonon field amplitude over time (see Fig.\ref{fig:anatomy}(a)).  Following some period of time, the coherence of the system is established, which results in an abrupt increase in the amplitude of the phonon field ($t = 260/\omega_m$). This leads to the appearance of a superradiant phase transition.  Moreover, an examination of the dynamics of the spin components depicted in Fig.\ref{fig:anatomy}(b) reveals the process of establishing the coherence of the system. This coherence is determined by the complete transition channel associated with the term $2\Omega S_x$ (without phonon) and the transition channel connected to the term  $i\frac{2G} {\sqrt{N}} (b_i - b_i ^\dagger) S_y$ (with phonon). As illustrated in Fig.\ref{fig:anatomy}(b), the term $2\Omega S_x$ is initially predominant. The collective spin rotates within the $X-Z$ plane of the Bloch sphere (where the averages of $X$ and $Z$ undergo change whilst the average of $Y$ remains constant prior to $t = 260 /\omega_m$). However, before this time, collective coherence is not established. Subsequently, after a certain duration ($t > 260 /\omega_m$), the spin aligns with the $X-Y$ plane (where the average of $Y$ undergoes a sudden change), collective coherence is established, and the term $i\frac{2G} {N} (b_i - b_i^\dagger) S_y$ becomes dominant. This results in the instantaneous occurrence of a superradiant phase transition. As demonstrated in Fig.\ref{fig:anatomy}(c) and (d), enhancing the spin-phonon coupling strength, $\lambda/\omega_m=0.8$, results in the rapid establishment of system coherence. Consequently, the emergence of the superradiant phase transition following the initiation of the driving light is characterized by a minimal delay.

\section{Quantum fluctuations and stability analysis}
\label{app:fluctuations}

To investigate the stability of the steady states and the divergence of 
fluctuations near the phase boundary, we extend the linearization of 
Appendix~\ref{app:lambda} to include input quantum noise. Adding the 
phonon vacuum noise $\delta\hat b_{\rm in}$ to the Heisenberg--Langevin 
equation, the fluctuations 
$\bm f=(\delta Q,\delta P,\delta X,\delta Y,\delta Z)^T$ obey
\begin{equation}
\dot{\bm f}=\mathbf A\,\bm f+\bm\eta(t),\qquad
\bm\eta=(\sqrt{2\kappa}\,\delta Q_{\rm in},\,\sqrt{2\kappa}\,\delta P_{\rm in},\,0,\,0,\,0)^T,
\label{eq:fluct_eq}
\end{equation}
where $\mathbf A$ is the same drift matrix as in Eq.~(\ref{eq:drift_full}), 
evaluated at the relevant steady state.

For a stable steady state (all eigenvalues of $\mathbf A$ satisfy 
$\mathrm{Re}\,\mu_j<0$), the symmetric covariance matrix $\mathbf V$ with 
elements $V_{ij}=\frac{1}{2}\langle f_if_j+f_jf_i\rangle$ obeys the 
Lyapunov equation
\begin{equation}
\mathbf A\,\mathbf V+\mathbf V\,\mathbf A^T=-\mathbf D,\qquad
\mathbf D=\mathrm{diag}(\kappa,\kappa,0,0,0),
\label{eq:lyapunov}
\end{equation}
where $\mathbf D$ is the diffusion matrix from the phonon vacuum noise. 
The phonon-number fluctuation follows as 
$\langle\delta\hat b^\dagger\delta\hat b\rangle=(V_{11}+V_{22}-1)/2$, 
which we evaluate numerically.

\begin{figure*}[t!]
\centering
\includegraphics[width=\textwidth]{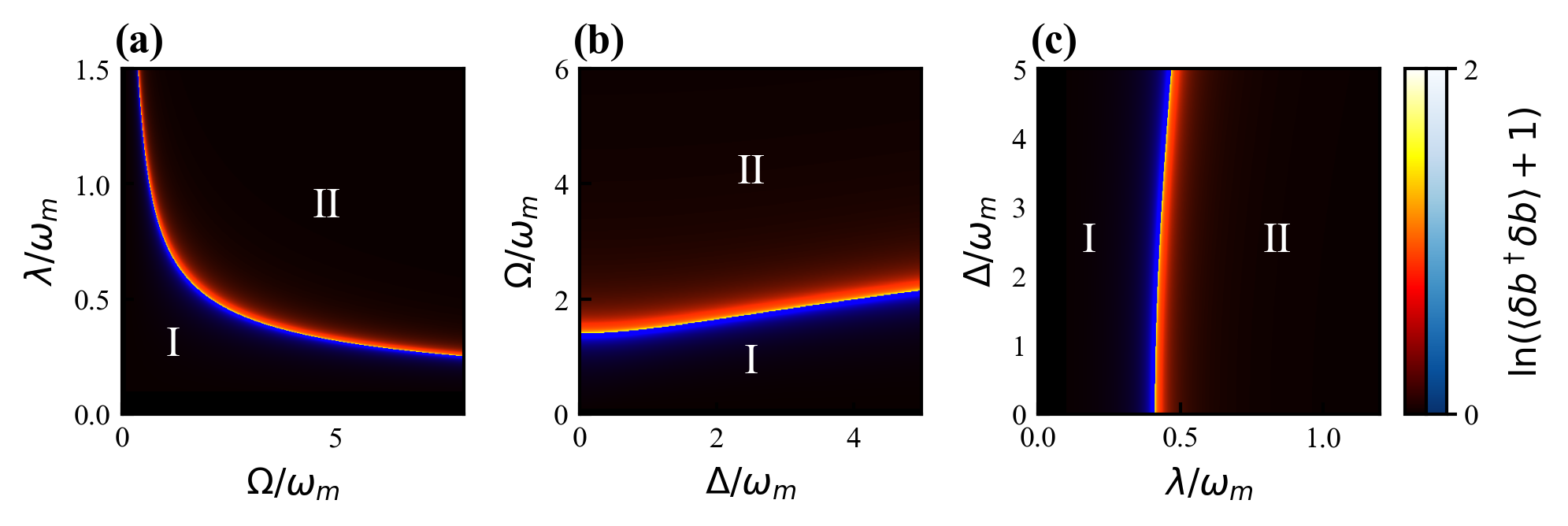}
\caption{Quantum fluctuation phase diagram ($N\to\infty$). Color: 
$\ln(\langle\delta\hat b^\dagger\delta\hat b\rangle+1)$; blue~=~NP, 
red~=~SP. The bright boundary indicates strongly enhanced fluctuations 
near the critical point. (a)~Fixed $\Delta/\omega_m=1.0$. 
(b)~Fixed $\lambda/\omega_m=0.6$. (c)~Fixed $\Omega/\omega_m=3.0$. 
Parameters: $\kappa=0.1\,\omega_m$.}
\label{fig:fluctuation}
\end{figure*}

Figure~\ref{fig:fluctuation} presents the fluctuation phase diagram in 
three parameter planes. The color encodes 
$\ln(\langle\delta\hat b^\dagger\delta\hat b\rangle+1)$: Region~I (NP) 
shows minimal fluctuations dominated by zero-point noise; Region~II (SP) 
shows finite fluctuations on top of the macroscopic phonon field; the 
bright boundary between them, where fluctuations are strongly enhanced, 
precisely matches $\lambda_c$ from Eq.~(\ref{eq:critical_coupling}). 
This enhancement reflects the increased susceptibility of the system to 
small perturbations near the critical point, a generic feature of 
continuous phase transition, i.e., second order phase transition.

\section{Finite-$N$ effects: theoretical derivation and detailed analysis}
\label{app:finiteN}

This appendix provides the theoretical underpinning for the finite-$N$ results presented in Sec.~\ref{sec:finiteN} of the main text. We derive the effective mixed-state form of the steady-state density matrix that produces the three-peak Wigner function near the phase boundary, derive the scaling laws for the peak separation and vacuum weight, and present the three-Gaussian fitting procedure used to extract these quantities from numerics.

\subsection{Origin of the three-peak Wigner structure}

For finite $N$, the steady-state phonon density matrix $\rho_{\rm ss}^{\rm ph}={\rm Tr}_{\rm spin}[\rho_{\rm ss}]$ does not collapse onto a single semiclassical solution.  With the light driving the NV centers, there are two transition channels between the ground state $|g\rangle$ and the excited state $|e\rangle$. It is clear that one of the channels is (the term $i\frac{2G} {\sqrt{N}} (b_i - b_i ^\dagger) S_y$) is associated with phonon emission, which is a prerequisite for determining the superradiant phase. In contrast, the other channel (the term $2\Omega S_x$), which does not involve phonon emission, is connected to the vacuum state. The two-channel configuration has been found to be competitive. In the NP region, the channel without phonon emission is the dominant one. Conversely, in the superradiant region, the channel accompanied by phonon emission has been demonstrated to be dominant. In the middle region, the coexistence of NP and SP has been observed, resulting in the emergence of a three-peak structure.
%Two physical mechanisms contribute simultaneously. (i)~The mean-field equations [Eq.~(\ref{eq:order_parameter})] admit two symmetry-broken steady states $b_0=\pm i|b_0|$ related by the $\mathbb{Z}_2$ symmetry $\hat{b}\to-\hat{b}$, $S_y\to -S_y$ of $H_{\rm eff}$. At any finite $N$, quantum fluctuations connect these two branches, so the steady state is an even mixture of both. (ii)~The drive term $2\Omega S_x$ in $H_{\rm eff}$ does not involve phonons and therefore does not directly populate the phonon mode; a fraction of the steady-state weight is retained near the vacuum, especially for small $N$ where the macroscopic occupation $N|b_0|^2$ is not yet large enough to suppress this vacuum component (see the scaling argument below).

The three-peak structure---the two superradiant branches and the residual vacuum---are captured by the mixed-state ansatz
\begin{equation}
\begin{aligned}
\rho_{\rm ss}^{\rm ph}&\approx p_0\,|0\rangle\!\langle 0|+\frac{1-p_0}{2}\bigl(|\alpha_+\rangle\!\langle\alpha_+|+|\alpha_-\rangle\!\langle\alpha_-|\bigr),\\
\alpha_\pm &= \pm i\sqrt{N}\,|b_0|,
\label{eq:rho_mixture}
\end{aligned}
\end{equation}
where $|0\rangle$ is the phonon vacuum and $|\alpha_\pm\rangle$ are coherent states displaced along the $\mathrm{Im}\,\alpha$ axis (the imaginary axis is selected by the $i(\hat{b}-\hat{b}^\dagger)S_y$ form of the spin--phonon coupling, which couples to phonon momentum rather than position). The corresponding Wigner function is
\begin{equation}
\begin{aligned}
    W(\alpha) &= p_0\,W_0(\alpha) + \frac{1-p_0}{2}\bigl[W_+(\alpha)+W_-(\alpha)\bigr],\\
    W_0(\alpha)&=\tfrac{2}{\pi}e^{-2|\alpha|^2},\\
    W_\pm(\alpha)&=\tfrac{2}{\pi}e^{-2|\alpha-\alpha_\pm|^2}.
    \label{eq:W_mixture}
\end{aligned}
\end{equation}
This form reproduces the three-peak structure observed numerically [insets of Fig.~\ref{fig:finiteN}(a)] and provides the basis for the fitting procedure below.

\subsection{Peak separation and the $\sqrt{N}$ scaling}

From Eq.~(\ref{eq:rho_mixture}), the phase-space separation between the two superradiant peaks is
\begin{equation}
    \Delta\alpha_{\rm im} = |\alpha_+ - \alpha_-| = 2\sqrt{N}\,|b_0|.
    \label{eq:peak_sep}
\end{equation}
The $\sqrt{N}$ scaling has a transparent origin: the rescaling $\langle\hat{b}\rangle=\sqrt{N}\,b_0$ used in Sec.~\ref{semi} maps the mean-field amplitude $b_0$ (of order unity in the SP) onto a coherent-state displacement $\sqrt{N}|b_0|$ of the bare phonon mode. Consequently, $\Delta\alpha_{\rm im}$ provides a direct phase-space measure of the order parameter $|b_0|$, and a linear fit of $\Delta\alpha_{\rm im}$ against $\sqrt{N}$ yields $|b_0|$ as the slope. Numerical verification is shown in Fig.~\ref{fig:finiteN_SM}(a): linear fits with $R^2>0.999$ confirm Eq.~(\ref{eq:peak_sep}) for all $\lambda$ studied, and the extracted slopes agree with the mean-field $|b_0|$ from Eq.~(\ref{eq:order_parameter}) within a few percent.

\subsection{Exponential scaling of the vacuum weight}

The vacuum weight $p_0$ in Eq.~(\ref{eq:rho_mixture}) reflects the competition between the two transition channels of $H_{\rm eff}$ [Eq.~(\ref{eq:H_eff})]: the phonon-mediated channel $i\frac{2G}{\sqrt{N}}(\hat{b}-\hat{b}^\dagger)S_y$, with effective matrix element $\sim G\sqrt{N}\,|b_0|=\lambda|b_0|\sqrt{N}/\omega_m\cdot\Omega$ in the macroscopically displaced subspace, and the direct optical drive $2\Omega S_x$, with matrix element $\sim\Omega$. The dimensionless ratio
\begin{equation}
    r \equiv \frac{\lambda|b_0|\sqrt{N}}{2\omega_m}
    \label{eq:ratio_r}
\end{equation}
controls which channel dominates: for $r\ll 1$ the optical drive prevails and the vacuum component survives, whereas for $r\gg 1$ the phonon-mediated channel dominates and the system is fully displaced into the SP. To quantify the competition, we note that the macroscopic energy difference between the vacuum and superradiant configurations grows as $\Delta E\sim\omega_m N|b_0|^2$ (the energy stored in $N|b_0|^2$ phonons); in the driven-dissipative steady state, the relative population of the higher-energy (vacuum) configuration is exponentially suppressed by an effective Boltzmann-like factor with an effective temperature set by the drive and dissipation, yielding
\begin{equation}
    p_0 \propto \exp\!\bigl(-\eta\, N\,|b_0|^2\bigr),
    \label{eq:p0_scaling}
\end{equation}
where $\eta$ is a dimensionless fitting parameter that depends on $\Omega,\Delta,\kappa$ and on the deviation of $|\alpha_\pm\rangle$ from ideal coherent states. Equation~(\ref{eq:p0_scaling}) implies two consequences: at fixed $\lambda$ (hence fixed $|b_0|^2$), $p_0$ decreases exponentially with $N$, so the three-peak structure disappears in the thermodynamic limit and the mean-field treatment becomes exact---consistent with the convergence shown in Fig.~\ref{fig:finiteN}(a); at fixed $N$, $p_0$ decreases exponentially with $|b_0|^2$, explaining why even for the smallest systems ($N=2$) deep in the SP ($\lambda/\omega_m=0.65$), the vacuum component is already strongly suppressed [Fig.~\ref{fig:finiteN_SM}(b)].

\subsection{Three-Gaussian fitting procedure}

\begin{figure*}[t!]
\centering
\includegraphics[width=\textwidth]{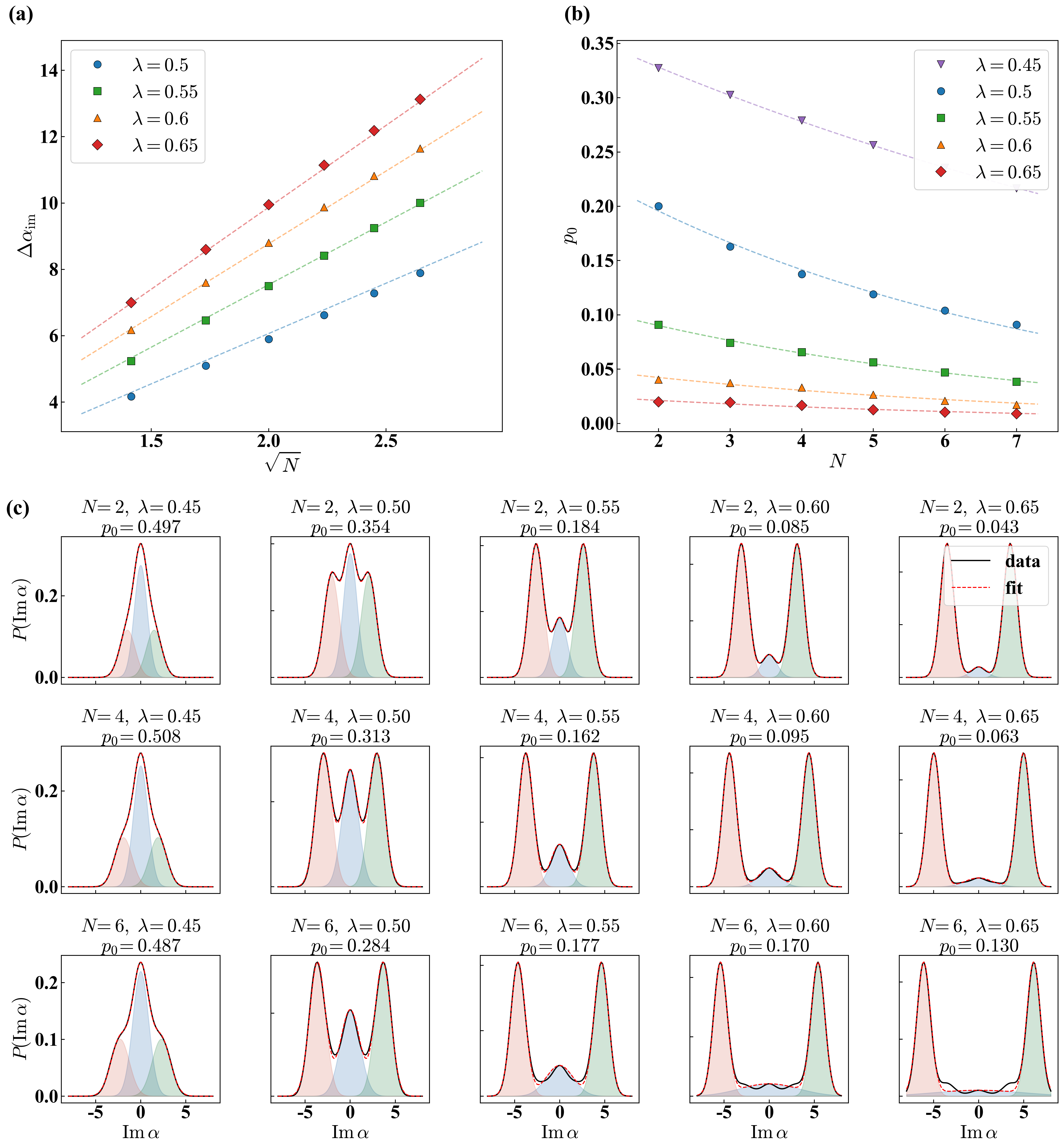}
\caption{Detailed finite-$N$ analysis. (a)~Peak separation $\Delta\alpha_{\rm im}$ versus $\sqrt{N}$ extracted from three-Gaussian fits; dashed lines are linear fits confirming the scaling of Eq.~(\ref{eq:peak_sep}) with $R^2>0.999$. (b)~Vacuum weight $p_0$ versus $N$ from the same fits; dashed curves are the exponential prediction $p_0=A_\lambda\exp(-\eta_\lambda N|b_0|^2)$ from Eq.~(\ref{eq:p0_scaling}). (c)~Three-Gaussian fit diagnostics for $N=2,4,7$ (rows) and $\lambda/\omega_m=0.45$--$0.65$ (columns). Black solid: numerical marginal $P(\mathrm{Im}\,\alpha)$; red dashed: three-Gaussian fit [Eq.~(\ref{eq:three_gauss})]; shaded: vacuum component (blue), $\alpha_+$ component (green), $\alpha_-$ component (red). Parameters: $\Delta/\omega_m=1$, $\Omega/\omega_m=3$, $\kappa=0.1\,\omega_m$.}
\label{fig:finiteN_SM}
\end{figure*}

To extract $p_0$ and $\Delta\alpha_{\rm im}$ quantitatively, we project the Wigner function onto the imaginary axis and fit the resulting marginal to a three-Gaussian model:
\begin{equation}
\begin{aligned}
    P(p) &= \int d(\mathrm{Re}\,\alpha)\,W(\mathrm{Re}\,\alpha+i p)\\
    &= p_0\,\mathcal{G}(p;0,\sigma_0) + \frac{1-p_0}{2}\bigl[\mathcal{G}(p;\alpha_{\rm fit},\sigma_1)+\mathcal{G}(p;-\alpha_{\rm fit},\sigma_1)\bigr],
    \label{eq:three_gauss}
\end{aligned}
\end{equation}
with $\mathcal{G}(p;\mu,\sigma)=(2\pi\sigma^2)^{-1/2}\exp[-(p-\mu)^2/(2\sigma^2)]$. The four free parameters $\{p_0,\alpha_{\rm fit},\sigma_0,\sigma_1\}$ are determined by least-squares fitting. Representative fits for $N=2,4,7$ across five coupling strengths $\lambda/\omega_m=0.45$--$0.65$ are shown in Fig.~\ref{fig:finiteN_SM}(c); residuals are below the percent level in all cases, confirming that the ansatz of Eq.~(\ref{eq:rho_mixture}) captures the essential physics of the finite-$N$ steady state.

\subsection{Implications for observable quantities}

Combining Eqs.~(\ref{eq:rho_mixture}) and (\ref{eq:p0_scaling}), the mean and variance of the phonon number per emitter become
\begin{equation}
\begin{aligned}
    \frac{\langle n\rangle}{N} &= (1-p_0)\,|b_0|^2,\\
    \mathrm{Var}(n) &\approx p_0(1-p_0)\,(N|b_0|^2)^2 + (1-p_0)\,N|b_0|^2,
    \label{eq:n_var}
\end{aligned}
\end{equation}
where the first term in $\mathrm{Var}(n)$ arises from the bimodal vacuum/superradiant occupation and the second from the intrinsic Poissonian fluctuations of the coherent states. The mean-value formula explains why the finite-$N$ curves in Fig.~\ref{fig:finiteN}(a) lie below the mean-field $|b_0|^2$ near $\lambda_c$, where $p_0$ is largest, and converge to the mean-field result deep in the SP, where $p_0\to 0$. Near the critical point the first term in the variance dominates, producing strongly super-Poissonian statistics $\mathrm{Var}(n)\gg\langle n\rangle$, which provides a direct experimental signature of the NP-SP coexistence accessible via phonon-number-resolved measurements.

\FloatBarrier
\twocolumngrid
\bibliography{my_references.bib}

\end{document}